\documentclass[aps,twocolumn,superscriptaddress,showpacs,amssymb,prl,floatfix]{revtex4-1}

\usepackage{graphicx,color}
\usepackage{dcolumn}
\usepackage{amsmath}
\usepackage{amssymb}
\usepackage{epstopdf}

\usepackage{hyperref}
\hypersetup{colorlinks=true,linkcolor=blue,
  implicit=true,breaklinks=true,pagebackref=true,backref=true,
  bookmarks=true,bookmarksnumbered=true,hyperfootnotes=true,debug=true,
  naturalnames=false,citecolor=blue,pdfview=FitH,pdfstartview=FitH,hyperindex=true}

\newcommand{\slrrtext}  {spin-lattice-relaxation rate}
\newcommand{\bacofeasx}     {Ba(Fe$_{1-x}$Co$_x$)$_2$As$_2$}

\newcommand{\slrr}      {$T_1^{-1}$}

\begin{document}

\thispagestyle{myheadings}

\title{Coexistence of cluster spin glass and superconductivity in Ba(Fe$_{1-x}$Co$_x$)$_2$As$_2$ for $0.060\leq x\leq 0.071$}

\author{A. P. Dioguardi}
\author{J. Crocker}
\author{A. C. Shockley}
\author{C. H. Lin}
\author{K. R. Shirer}
\author{D. M. Nisson}
\author{M. Lawson}
\author{N. apRoberts-Warren}
\affiliation{Department of Physics, University of California, Davis, California 95616, USA}
\author{P. C. Canfield}
\author{S. L. Bud'ko}
\author{S. Ran}
\affiliation{Ames Laboratory U.S. DOE and Department of Physics and Astronomy, Iowa State University, Ames, Iowa 50011, USA}
\author{N. J. Curro}
\affiliation{Department of Physics, University of California, Davis, California 95616, USA}
\email{dioguardi@ms.physics.ucdavis.edu}

\date{\today}

\begin{abstract}
We present $^{75}$As nuclear magnetic resonance data from measurements of a series of Ba(Fe$_{1-x}$Co$_x$)$_2$As$_2$ crystals with $0.00 \leq x \leq 0.075$ that reveals the coexistence of {frozen antiferromagnetic domains and superconductivity for $0.060 \leq x \leq 0.071$.  Although bulk probes reveal no long range antiferromagnetic order beyond $x=0.06$, we find that the local spin dynamics reveal no qualitative change across this transition. }The characteristic domain sizes vary by more than an order of magnitude, reaching a maximum variation at $x=0.06$. This inhomogeneous glassy dynamics may be an intrinsic response to the competition between superconductivity and antiferromagnetism in this system.

\end{abstract}


\maketitle

The iron arsenide family of superconductors have reignited interest in the physics of strongly correlated electron systems since the discovery of superconductivity below 43 K in LaFeAsO$_{1-x}$F$_x$ in 2007 \cite{LaOFFeAsNature,johrendtPRL2008,LiFeAsdiscovery,KFe2Se2Discovery}.
Some of the most widely investigated materials are the electron-doped  Ba(Fe$_{1-x}$T$_x$)$_2$As$_2$ (T = Co, Ni) because not only are large single crystals available but the phase diagrams exhibit a rich interplay of order parameters \cite{doping122review,JohnstonPnictideReview}.  A key feature of this system is the continuous suppression of the antiferromagnetic order giving rise to a putative quantum phase transition below a dome of superconductivity, similar to the cuprate and heavy fermion superconductors \cite{imaiBa122overdoped,FernandesNS122,NakaiPdopedBa122PRL,tuson}. Competing orders often emerge at a quantum critical point, and long range interactions can give rise to intrinsic inhomogeneity \cite{KivelsonPhaseSeparation,Curro2000b,demler,*SachdevCompeteCupratesPRB2009}. In some superconducting families these competing orders coexist microscopically \cite{kitaokaRh115pressure,Urbano2007,Lefebvre:2000vn}, whereas in others the electronic degrees of freedom become inhomogeneous on mesoscopic length scales \cite{YazdaniLocalOrderPGScience2004}.  Reports in the electron doped BaFe$_2$As$_2$ system have been mixed: some studies indicate  homogeneous coexistence of antiferromagnetism and superconductivity \cite{Laplace2009,JulienDopedIronArsenides,CanfieldCoexistenceBa122,IshidaPdopedBa122JPSJ2012,NiCanfield122review}, however others suggest that these two orders do not coexist microscopically \cite{DaiCoexistencePRLNi122,DaiNature2012,muSRBaCo122}.
Here we report $^{75}$As nuclear magnetic resonance (NMR) data that unambiguously reveal an {inhomogeneous distribution of frozen antiferromagnetic domains in the superconducting state of \bacofeasx.  Our results suggest that such cluster spin glass phases may be a general feature of materials in which superconductivity emerges upon doping an antiferromagnetic system.}

\begin{figure*}
\includegraphics[width=\linewidth]{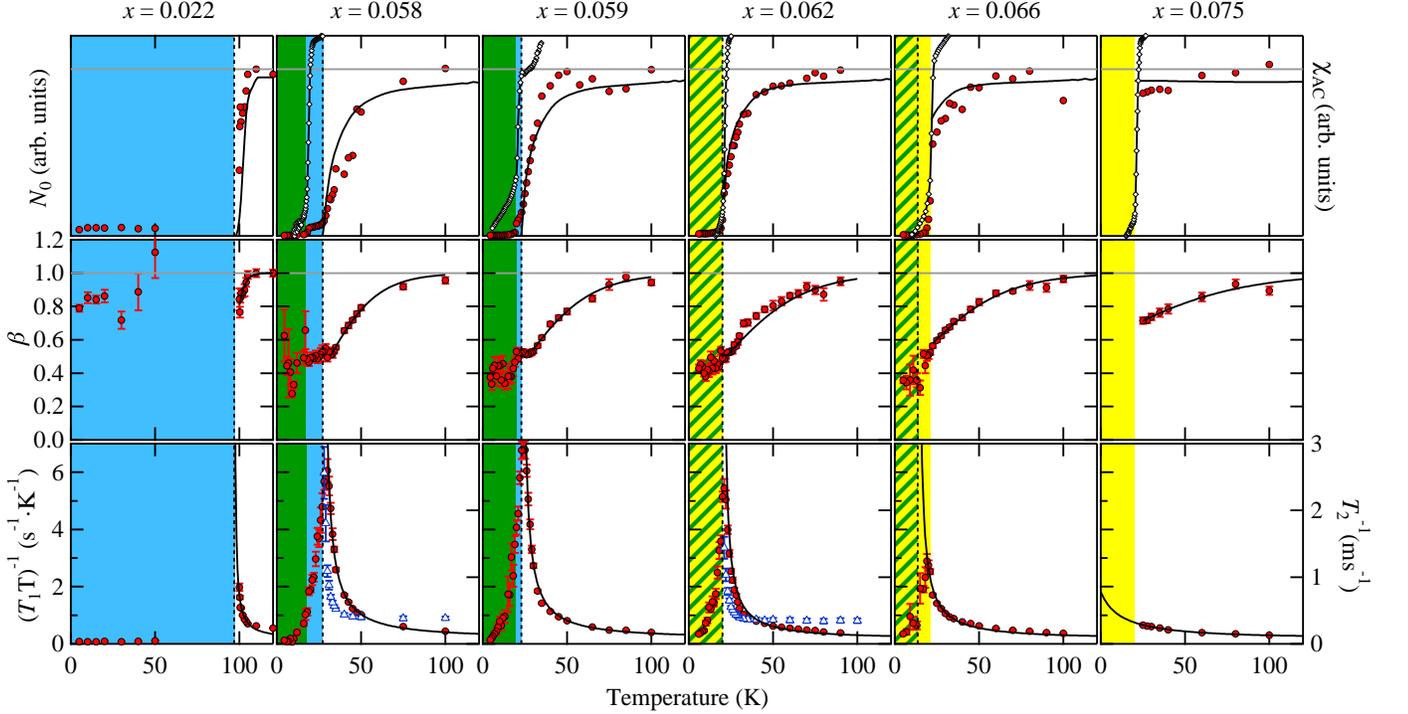}
\caption{(Color Online) The signal intensity ($N_0$, upper row), stretching exponent ($\beta$, middle row) and $(T_1T)^{-1}$ (lower row) versus temperature for several different dopings for $H_0\perp c$ (data for $H_0 \parallel c$ is provided in the supplemental information).  $\chi_{AC}$ ($\lozenge$, upper row) and $T_2^{-1}$ ($\vartriangle$, lower row) are shown on the  right hand axes. All the data were acquired in 11.7 T except for the $x=0.022$ sample measured in 8.5 T. Blue (yellow) shading indicates bulk antiferromagnetism (superconductivity), green indicates bulk coexistence, and the hatched pattern indicates the frozen cluster glass regime. The solid lines in the middle and lower rows are fits to $\beta$ and $(T_1T)^{-1}$ versus temperature, and in the upper row the solid lines are simulations of $N_0$ as discussed in the text.The dotted vertical lines indicate $\theta$.
\label{fig:gridsummary}}
\end{figure*}

Single crystals of \bacofeasx\ were grown by the self-flux method and the Co concentrations were determined via microprobe analysis \cite{doping122review,NiCanfield122review}. $^{75}$As (100\% abundant, $I=3/2$) NMR spectra, \slrrtext, \slrr, and spin echo decay rates, $T_2^{-1}$, were measured in fixed fields of 8.75 T and  11.7 T by acquiring spin echoes using standard pulse sequences for a range of Co concentrations between 0.00 and 0.075. Local variations in the ordered Fe/Co moments near dopants broaden the NMR spectra for the field oriented along the (001) direction \cite{Dioguardi2010}. Because this broadening significantly reduces the signal intensity we focus primarily  on measurements  with the field oriented in the $ab$ plane {(spectra are available in the online supplemental information). Our samples exhibit sharp NMR spectra (FWHM 20-30 kHz) and microprobe analysis indicates a standard deviation of less than 5\% of the doping, indicating high sample quality and homogeneity.}   The superconducting transition temperature $T_c$ was measured \textit{in situ} using the NMR coil in field, and  the AC susceptibility is shown in the upper series of panels in Fig. \ref{fig:gridsummary}.

\begin{figure}
	\centering
\includegraphics[width=0.95\linewidth]{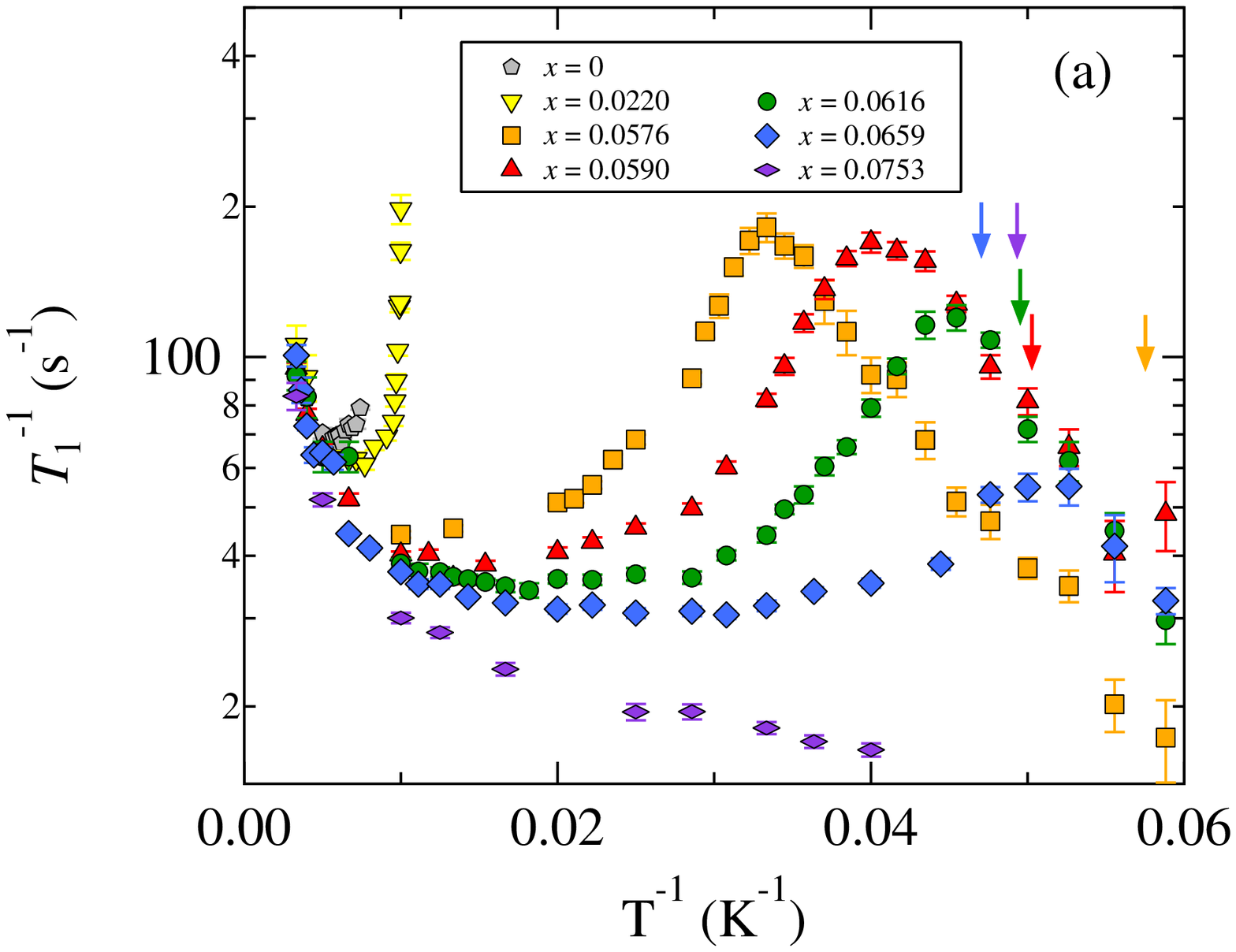}
\includegraphics[width=0.95\linewidth]{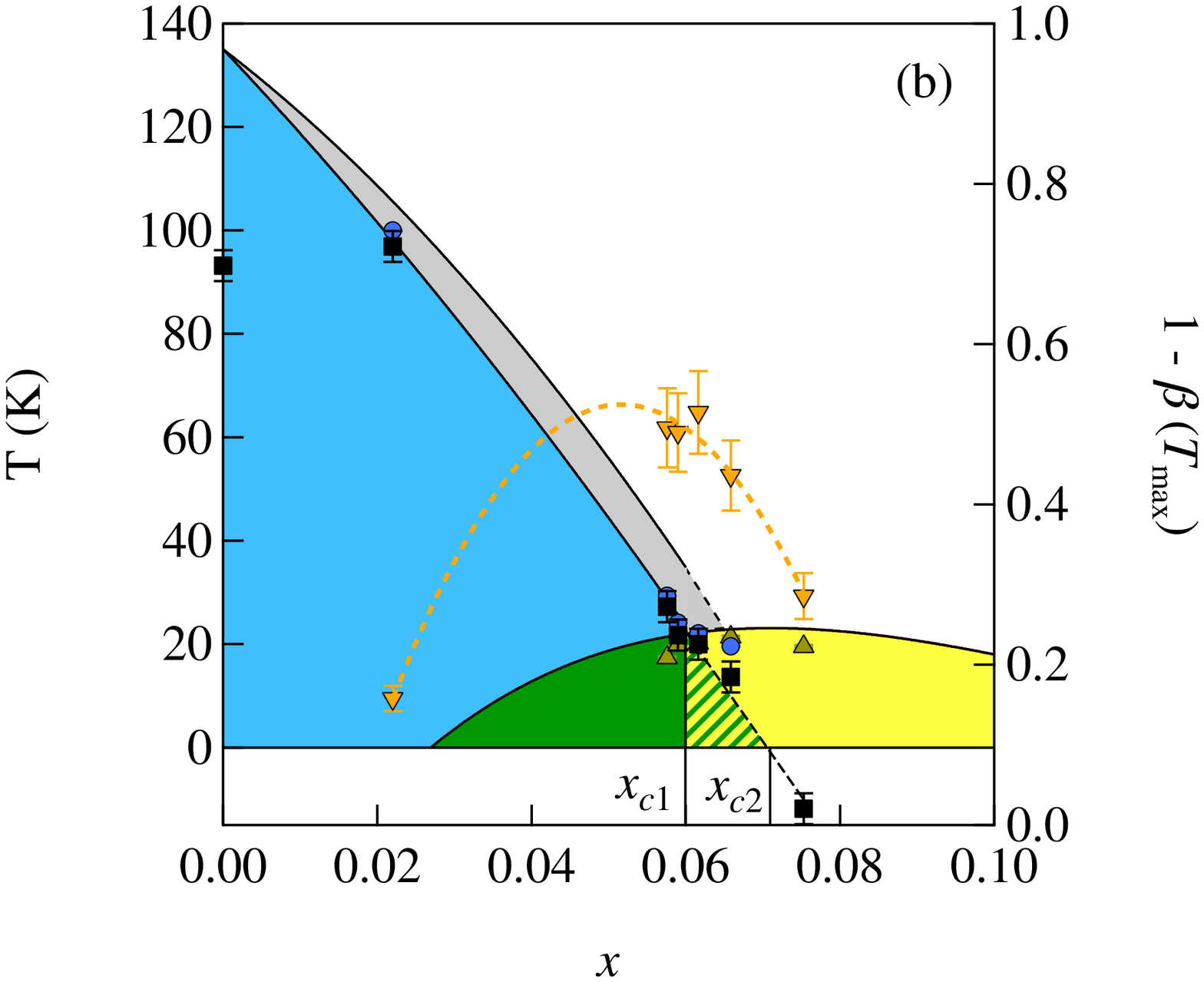}
\caption{\label{fig:phasediagram}(Color Online) (a) \slrr\ vs. $T^{-1}$; arrows indicate $T_c$. (b) Phase diagram for \bacofeasx: solid black lines are polynomial fits to the bulk phase diagram in zero field \cite{NSBa122incommensurate,FisherBa122PhasediagramPRB2009,Ba122Hc2study} indicating antiferromagnetic order (blue), orthorhombic structure (gray), superconductivity (yellow), coexisting superconductivity and antiferromagnetism (green), and cluster glass (hatched pattern).  The data points are $\theta$ ($\blacksquare$), $T_{\rm max}$ ($\bullet$), $T_c$ ($\blacktriangle$) at 11.7 T, and $1-\beta(T_{\rm max})$ ($\blacktriangledown$, right axis).  The dashed black line is a  linear fit to $\theta$ and the dashed orange line is a guide to the eye  for $1-\beta(T_{\rm max})$, which is a measure of the degree of inhomogeneity.}
\end{figure}

The \slrrtext\ was measured by inversion recovery at the central transition ($I_z = +\frac{1}{2} \leftrightarrow -\frac{1}{2}$), and the data are summarized in Figs. \ref{fig:gridsummary} and \ref{fig:phasediagram}(a).  The data are well fit by the standard normal-modes recovery function for a spin 3/2 nucleus for temperatures $T \gtrsim 100$ K \cite{Narathrecovery}, but below this temperature the quality of fit decreases significantly.  To account for this change of behavior we fit the data to a stretched exponential form \cite{BorsaPRL1993}:
\begin{equation}
	M(t) = M_0\left[1\!-\!\!2f\left(\frac{9}{10}e^{-(6t/T_1)^\beta}\!\!+ \frac{1}{10}e^{-(t/T_1)^\beta}\right)\right],
	\label{eqn:spin_3over2_strexp}
\end{equation}
where $M_0$ is the equilibrium nuclear magnetization, $f$ is the inversion fraction and $\beta$ is the stretching exponent. A fit with $\beta<1$ indicates a distribution of relaxation rates, and we find that $M_0$, $\beta$, and $T_1$ are each strong functions of both temperature and doping.

The bottom series of panels in Fig. \ref{fig:gridsummary} shows $(T_1T)^{-1}$ versus temperature and doping.  Our data are consistent with  published data \cite{imaiBa122overdoped,NakaiPdopedBa122PRL,IshidaPdopedBa122JPSJ2012,NakaiPRB2013}, but extend to lower temperatures than reported previously revealing a peak at $T_{\rm max}$ that coincides with bulk measurements of the N\'{e}el temperature, $T_N$, for $x < x_{c1} = 0.06$. Following Ning \textit{et al.} \cite{imaiBa122overdoped} we fit the temperature dependence to the sum of three terms: $(T_1T)^{-1} = \left(a + b e^{-\Delta/k_B T}\right) + c/(T - \theta)$, where the first two terms represent intra-band spin fluctuations with  $a = 0.11$ s$^{-1}$K$^{-1}$, $b = 0.63$  s$^{-1}$K$^{-1}$ and $\Delta = 450$ K, and the last term is doping dependent and represents inter-band antiferromagnetic spin fluctuations. Fig. \ref{fig:phasediagram}(b) shows $\theta$ and $T_{\rm max}$ as a function of doping.  Previous NMR and neutron scattering experiments suggested that $x_{c1}$ corresponds to a quantum critical point (QCP) where long-range antiferromagnetism disappeared and the nature of the spin fluctuations changed character  \cite{imaiBa122overdoped,CanfieldCoexistenceBa122}. As seen in Fig. \ref{fig:phasediagram}(b) $T_N$ intercepts $T_c$ at $x_{c1}$
and long range antiferromagnetism is absent for $x>x_{c1}$.
\emph{Surprisingly, our data reveal no qualitative change across this boundary, in contrast to the response that would be expected at a QCP.  We find that $\theta$ does not vanish at $x_{c1}$, but continues to remain finite and extrapolates smoothly from lower dopings.} $(T_1T)^{-1}$
continues to diverge at $T_{\rm max}$, except that for $x>x_{c1}$ $T_{\rm max}$ coincides with $T_c$ whereas $\theta< T_{\rm max}$ and uniformly extrapolates to $T=0$ at $x_{c2}=0.071\pm0.003$. This result indicates that \emph{for $x_{c1}\leq x \leq x_{c2}$} some of the As sites are probing slow antiferromagnetic fluctuations,
a result that is  consistent with recent muon spin rotation ($\mu$SR) observations \cite{muSRBaCo122}.
The drop in $(T_1T)^{-1}$ below $T_{\rm max}$ arises because the spin fluctuations in the ordered state change character. For $x<x_{c1}$, this suppression occurs because long range antiferromagnetic order develops prior to the onset of superconductivity; for $x>x_{c1}$ it is likely that the superconducting condensate dampens the  spin fluctuations and cuts short the development of long range order antiferromagnetism.

Evidence for inhomogeneity is seen in the middle series of panels in Fig. \ref{fig:gridsummary} which show that the exponent $\beta$ decreases from unity at high temperature to a minimum of $\beta\sim 0.4$. Stretched exponential decay is often used to describe relaxation in disordered systems arising from a distribution, $\rho_{\beta}(W)$, of relaxation rates, $W$. For $\beta = 1$ $\rho_{\beta}(W)$ is a delta function centered at $T_1^{-1}$; as $\beta$ decreases this distribution broadens out over a range of $W$.  In this case  \slrr\ is  the median of the distribution, and $\beta$ is approximately equivalent to the logarithmic full width half maximum of $\rho_{\beta}(W)$ \cite{johnstonstretched}. A value of $\beta = 0.4$ indicates that the distribution of relaxation rates varies by more than an order of magnitude.  In the paramagnetic phase the temperature dependence of $\beta$ is well-fit to the expression $\beta(T) = \beta_0 + (1-\beta_0)\tanh[(T-\theta)/T^*]$, where $\beta_0$ and $T^*$ are parameters that represent the $T=0$ value and the temperature scale for the onset of the inhomogeneous distribution, respectively.  The quantity $1-\beta(T_{\rm max})$, shown in Fig. \ref{fig:phasediagram}(b), is a measure of the width of the distribution. This data indicates that the distribution of relaxation rates becomes broadest at $T_{\rm max}$ at the doping level $x_{c1}$.

Further evidence for spatial inhomogeneity is found in the upper series of panels in Fig. \ref{fig:gridsummary}.  In Eq. \ref{eqn:spin_3over2_strexp} the equilibrium magnetization $M_0\sim N_0 H_0/T$, where $H_0$ is the applied field and $N_0$ is the number of nuclei contributing to the NMR signal. In principle $N_0$ should be temperature independent, however there is a clear suppression of $N_0$ that develops below $\sim 70$K. This phenomenon, known as signal wipeout, indicates that not all of the nuclei are contributing to the NMR signal, and is often found in glassy systems such as the underdoped cuprates in which the characteristic spin fluctuation times can slow dramatically \cite{Curro2000b,JulienGlassyStripes2001PRB,singer}.  For a randomly fluctuating hyperfine field, $h(t)$, with autocorrelation time $\tau_c$, the spin echo intensity is proportional to $e^{-t/T_2}$, where $t$ is time since the excitation pulse and  $T_2^{-1} = \gamma^2 h_{\parallel}^2\tau_c$, where $h_{\parallel}^2$ is the mean squared hyperfine field parallel to $H_0$.  When $\tau_c$ is sufficiently large ($t/T_2\gg 1$) the spin echo can decay faster than the effective time window of the NMR spectrometer and the signal intensity is suppressed. This suppression can happen close to a magnetic phase transition, as well as in spin glasses where critical fluctuations exhibit long correlation times.

The wipeout can be understood quantitatively as a consequence of the increasing width of the distribution of relaxation rates, $\rho_{\beta}(W)$.  This distribution reflects a distribution of correlation times, $\tau_c$, such that $\rho_{\beta}(W) dW = P_{\beta}(\tau_c) d\tau_c$, where $W = \gamma^2 h_{\perp}^2\tau_c/(1+\omega_L^2\tau_c^2)$, $h_{\perp}$ is the perpendicular component of the fluctuating hyperfine field, and $\omega_L=\gamma H_0$ is the Larmor frequency. The NMR signal intensity is proportional to $N(T) =\int_0^{\tau_{cut}} P_{\beta}(\tau_c) d\tau_c = \int_0^{W_{cut}} \rho_{\beta}(W) dW$, where $\tau_{cut} \approx \ln(100)/2\gamma^2h_c^2t_{min}$, and $t_{min}\approx 10\mu$s.  For high temperatures where $\beta\approx 1$ most of the weight in $\rho_{\beta}(W)$ is below $W_{cut}$ and all of the nuclei contribute to the signal; as the distribution widens at low temperature and \slrr\ increases a greater fraction of the weight extends above $W_{cut}$ and therefore does not contribute to the signal.  $\rho_{\beta}(W)$ is given as an inverse Laplace transform, and can be computed numerically \cite{BerberanSantos2005171}. We compute $N(T)$ for each doping using the fitted values for $\beta(T)$ and $(T_1T)^{-1}$ and the results are compared with the measured data in the upper panels of Fig. \ref{fig:gridsummary}. Details of the calculation are provided in the online supplemental information.  The simulated wipeout clearly exhibits similar trends to the data, indicating that both the wipeout and the stretched behavior arise from an inhomogeneous distribution of local correlation times, $\tau_c$.  {Variations in $h_{\perp}$ and in doping can also be a source of \slrr\ inhomogeneity \cite{JulienDopedIronArsenides,NingJPSJ2009}, but the small values of $\beta$ and the wipeout suggest that $\tau_c$ variations are dominant.}

{
The inhomogeneity we observe for the $x=0.058$ and 0.059 samples can be explained by the presence of short-range incommensurate magnetic order as observed in neutron scattering \cite{NSBa122incommensurate,DaiCoexistencePRLNi122}.  However for $x>x_{c1}$ neutron scattering experiments reveal no long range order whereas NMR clearly indicates the presence of frozen moments. Taken together with the evidence for inhomogeneity observed in the wipeout and the stretched exponential behavior, these results suggest that local antiferromagnetic fluctuations develop in disconnected spatial regions with a distribution of domain sizes. These domains do not appear to be mobile correlated patches of short-range antiferromagnetism, which would result in a uniform relaxation rate {for all nuclei, as is typical for antiferromagnetic transitions in homogeneous systems \cite{Curro2000a}. Furthermore, } they do not exhibit activated or Vogel-Fulcher dynamics (see. Fig. \ref{fig:phasediagram}(a)) as observed in the lightly doped cuprates \cite{BorsaPRL1993,Curro2000b} and recently in  LaFeAsO$_{1-x}$F$_x$  \cite{Hammerath2013}.  In fact our data suggest that local antiferromagnetic fluctuations develop in disconnected spatial regions characteristic of cluster spin glass behavior, similar to that observed in underdoped superconducting cuprates \cite{JulienClusterGlassCuprates,MitrovicGlassy214PRB2008,BaekYBCOunderdopedPRB2012,JulienFreezingChargeOrderYBCOPRB2013}.  Within each domain the local correlation time $\tau_c\sim \xi^{z}$ where $\xi$ is the local correlation length and $z =1$ is the dynamical scaling exponent.
If  the domains are disconnected then when $\xi$ is greater than the domain size $\tau_c$ will saturate.  As a result, a distribution of domain sizes gives rise to a distribution of saturated $\tau_c$ and hence a distribution $\rho_{\beta}(W)$.  This distribution, therefore, reflects the properties of the domains, which must have a size distribution that varies by more than an order of magnitude in order to explain the NMR results.
}

{A key property of this distribution is that neighboring domains do not merge with one another as the correlation length grows, but remain disconnected.
In  superconducting La$_{1.94}$Sr$_{0.06}$CuO$_4$ frozen antiferromagnetic clusters of varying sizes are believed to be surrounded by hole-rich regions that reduce the magnetic coupling between domains \cite{JulienClusterGlassCuprates}.  This electronic inhomogeneity {may be} a consequence of long range interactions between doped holes in a Mott insulator \cite{KivelsonPhaseSeparation}.  It is unlikely, however, that charge inhomogeneity is the origin of the behavior we observe in \bacofeasx.  Rather, disorder in the local structure driven by the Co atoms and the presence of twin or 180$^{\circ}$ phase boundaries may give rise to a random distribution of domain sizes. We conjecture that superconductivity emerges in regions between the domains effectively disconnecting any correlations between the domains, cutting off the growth of antiferromagnetic fluctuations, and preventing the development of long range order.  It is natural for a competing order parameter to emerge in regions where the primary order is reduced, for example between domains \cite{GLmagnetioelectrictheory,Xiao2012} or at vortex cores in the mixed state \cite{HoffmanVortexCoresScience2002,AFMinVortexMitrovicPRB2003,Wu2013}.
}

In conclusion we find evidence for a cluster spin-glass coexisting with superconductivity in \bacofeasx\ for $x_{c1}\leq x \leq x_{c2}$ rather than a uniform crossover expected at a second order quantum phase transition.
{Although the critical concentration $x_{c1}$ corresponds to the absence of long-range antiferromagnetism, local order persists up to a QCP at $x_{c2}$, similarly to CeRhIn$_5$ under pressure \cite{tuson}.}
It is surprising that such inhomogeneous slow dynamics should emerge in  conducting {pnictide} systems since there is no evidence for electronic phase inhomogeneity as observed in the cuprates and other doped Mott insulators \cite{WestfahlStripeGlassPRB2001}.  It is possible the glassy behavior we observe arises either because of frustration generated by doping-induced disorder, or as an intrinsic consequence of competing orders in a nominally pure system \cite{NussinovPRB2009}.

We  thank H.-J. Grafe, F. Hammerath, P. Klavins, M. Graf, {J.-X. Zhu} and A. Balatsky for stimulating discussions and A. Thaler for assistance with initial sample growth. Work at UC Davis was supported by  the NSF under Grant No.\ DMR-1005393, and part of this  work  performed at the Ames Laboratory (PCC, SLB, SR) was supported by the U.S. Department of Energy, Office of Basic Energy Science, Division of Materials Sciences and Engineering. Ames Laboratory is operated for the U.S. Department of Energy by Iowa State University under Contract No. DE-AC02-07CH11358.

\bibliography{CurroBibliography}

\pagebreak

\section{Supplemental Information}

\begin{figure*}
	\centering
	\includegraphics[width=\linewidth]{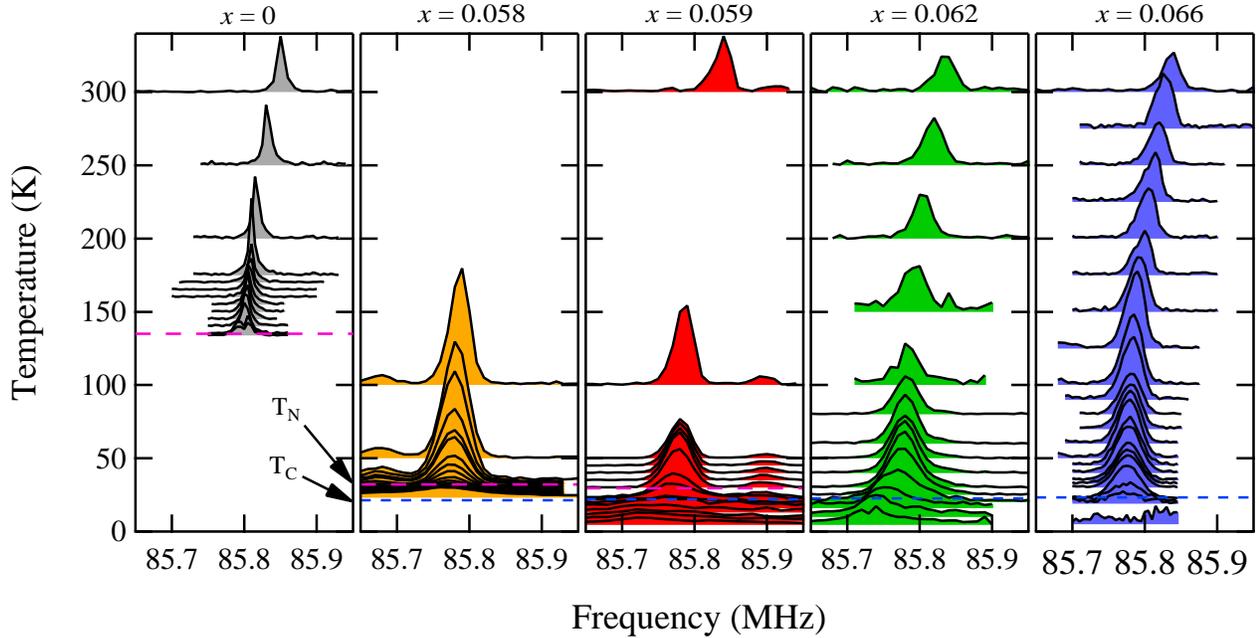}
	\caption{\label{fig:spectra_vs_temp}(Color Online) NMR frequency spectra as a function of temperature and doping for $x =$ 0, 0.058, 0.059, 0.062, and 0.066 in 11.7 T for field $\mathbf{H}_0\perp\hat{c}$. Pink horizontal dashed lines indicate $T_N(x)$ (from NS data~\cite{Pratt:2011cv}) and blue horizontal dashed lines indicate $T_C(x)$ as measured by \textit{in-situ} $\chi_{AC}$ measurements.}
\end{figure*}

\subsection{Frequency Swept Spectra for $H_0 \perp c$}

NMR frequency swept spectra were collected as a function of temperature with $H_0 \perp c$ at the central resonance ($I_z = +\frac{1}{2} \leftrightarrow -\frac{1}{2}$). These spectra were produced by integrating spin echoes and sweeping the excitation frequency. Figure~\ref{fig:spectra_vs_temp} shows the results for $x =$ 0, 0.058, 0.059, 0.062, and 0.066. The $x = 0$ resonance has a Lorentzian line shape and shifts down in frequency of frequency. We observe a small splitting at $T_N$ but did not acquire data at lower temperatures. In the Co-doped samples we observe a Gaussian line shape and  a decrease in frequency with decreasing temperature. The spectral intensity in the doped samples is wiped out starting just above $T_N$ and $T_c$ as described in the main text. In the samples where spectra were collected below $T_N$ and $T_c$ the negative shift is increased and the lines broaden significantly.


\subsection{Spin-Lattice Relaxation for $H_0 \parallel c$}

\begin{figure}
\includegraphics[width=\linewidth]{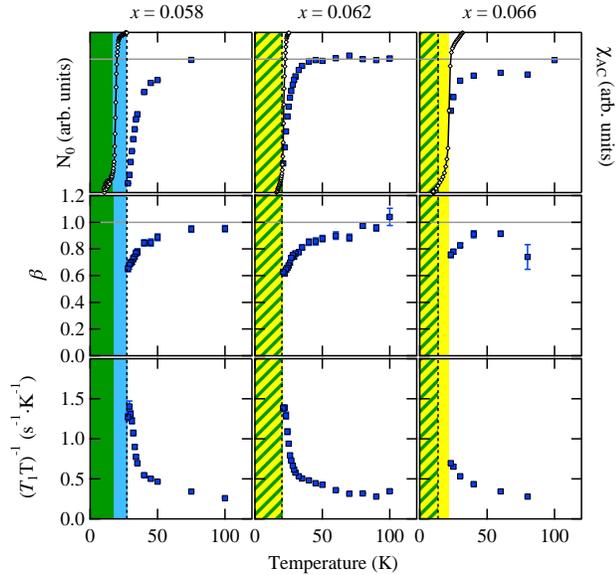}
\caption{(Color Online) The signal intensity ($N_0$, upper row), stretching exponent ($\beta$, middle row) and $(T_1T)^{-1}$ (lower row) versus temperature for several different dopings for $H_0\parallel c$.  $\chi_{AC}$ ($\lozenge$, upper row) is shown on the right hand axes. The $x = 0.058$ and $x = 0.062$ data were acquired in 8.794 T and 8.795 T respectively, and the $x = 0.066$ data were acquired in 3 T. Blue (yellow) shading indicates bulk antiferromagnetism (superconductivity), green indicates bulk coexistence, and the hatched pattern indicates the frozen cluster glass regime. 
\label{fig:gridsummarypar}}
\end{figure}

Data collected for $H_0 \parallel c$ is shown in Fig. \ref{fig:gridsummarypar}, and is qualitatively similar to that for the perpendicular direction.

\subsection{Wipeout Calculations}

\begin{figure}
	\centering
	\includegraphics[width=\linewidth]{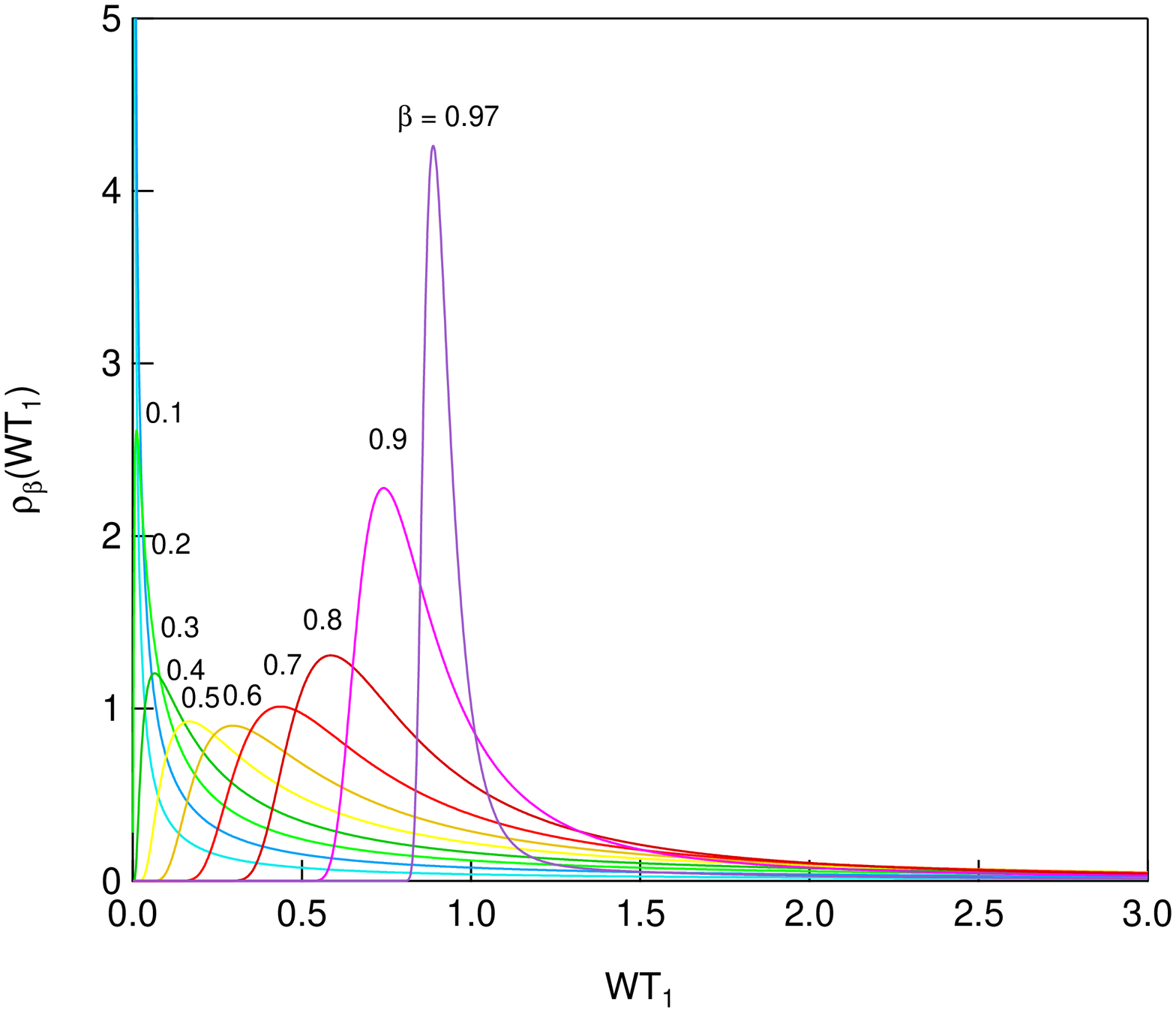}
	\caption{\label{fig:distribution}(Color Online) $\rho_{\beta}(WT_1)$ for various values of $\beta$. As $\beta\rightarrow 1$ the distribution approaches a delta function centered at $WT_1 = 1$.}
\end{figure}

The fraction of nuclei that contribute to the NMR signal is determined by the integral over the distribution $\rho_{\beta}(W)$ up to the cutoff value $W_{cut}$:
\begin{equation}
\nonumber N(T) = \int_0^{W_{cut}} \rho_{\beta}(W) dW = \int_0^{y_{cut}} \rho_{\beta}(y)dy,
\end{equation}
where $\beta(T)$  and $(T_1T)^{-1}$ are  given in the main text, and $y_{cut} = W_{cut}T_1$.
The distribution $\rho_{\beta}(W)$ can be approximated by:
\begin{widetext}
\begin{equation}
\nonumber\rho_{\beta}(W) = T_1 \frac{B}{(WT_1)^{(1-\beta/2)/(1-\beta)}}\exp\left[-\frac{(1-\beta)\beta^{\beta/(1-\beta)}}{{(WT_1)^{\beta/(1-\beta)}}} \right]\times
 \begin{cases}
   \frac{1}{1+C(WT_1)^{\beta(0.5-\beta)/(1-\beta)}} & \text{if } \beta \leq 0.5 \\
   {1+C(WT_1)^{\beta(\beta-0.5)/(1-\beta)} }      & \text{if } \beta > 0.5
  \end{cases},
 \end{equation}
\end{widetext}
where $B$ and $C$ are constants that are given in Table 1 of~\cite{BerberanSantos:2005jn}. (Note that for there is a typo in this table for $\beta = 0.5$, in which case $B = 0.282$.) This distribution is shown in Fig. \ref{fig:distribution} for several values of $\beta$.    The solid black lines shown in the upper row of Fig. 1 of the main text were calculated using this expression using the fitted parameters for $\beta$ and $(T_1T)^{-1}$, and using $W_{cut} = 95$ s$^{-1}$.  In the limit $\omega_L\tau_{cut} < 1$,  $W_{cut} = \gamma^2 h_{\perp}^2 \tau_{cut} \approx \ln(100)(h_{\perp}/h_c)^2/2t_{min}$, which implies that $h_{c}/h_{\perp}\sim 70$.  This ratio is in agreement with theoretical estimates of the the hyperfine fields based upon the form factors and the spin fluctuations that are peaked near the ordering wavevector $\mathbf{Q} = (\pi/a,0,\pi/c)$~\cite{Smerald:2011eq}.

\end{document}